\documentclass[10pt, twocolumn, comsoc]{IEEEtran}

\usepackage{graphicx,epsfig}
\usepackage[noadjust]{cite}
\usepackage{mcite}
\usepackage{amsfonts,helvet}
\usepackage{fancyhdr}
\usepackage{threeparttable}
\usepackage{epsf,epsfig}
\usepackage{amsthm}
\usepackage{amsmath}
\usepackage{siunitx}
\usepackage{amssymb}
\usepackage{dsfont}
\usepackage{subfigure}
\usepackage{color}
\usepackage[linesnumbered,ruled]{algorithm2e}
\usepackage{enumerate}
\usepackage{cancel}
\usepackage{mathptmx}
\usepackage{comment}
\usepackage{siunitx}

\usepackage[normalem]{ulem}

\newtheorem{theorem}{Theorem}

\newtheorem{lemma}{Lemma}

\SetKwInput{KwInput}{Initialize}
\SetKwInput{KwOutput}{Output}
\SetKwProg{Fn}{Stage 1}{}{}
\SetKwProg{Fn}{Stage 2}{}{}

\usepackage{eucal}

\begin{document}

\title{
Asymptotic Scaling Law Analysis of Multicast Satellite Communications with Massive MIMO
}

\author{Seyong~Kim and Jeonghun~Park

\thanks{
This work was supported by Samsung Research Funding \& Incubation Center of Samsung Electronics under Project Number SRFC-IT2402-06.
S. Kim and J. Park are with the School of Electrical and Electronic Engineering, Yonsei University, Seoul 03722, South Korea (e-mail: {\texttt{sykim@yonsei.ac.kr; jhpark@yonsei.ac.kr}}). 
}} 

\maketitle \setcounter{page}{1} 

\begin{abstract} 
In this paper, we consider a geostationary orbit (GEO) satellite communication system that employs massive multiple-input multiple-output (MIMO) for multicast transmission. 
By modeling the spatial distribution of ground users using a Poisson point process (PPP) and assuming a fixed-beam precoding is adopted, we find a closed-form expression for the asymptotical rate scaling law as a function of the number of antennas and the scaling factors of user density and multicast users. 
From the derived analytical expression, we reveal that the rate degradation caused by multicast transmission can be precisely compensated by increasing the user density accordingly.

\end{abstract}


\section{Introduction}
To compensate for significant signal attenuation in satellite communications, using multiple-input multiple-output (MIMO) technology in satellites has been considered an appealing approach \cite{harris:milcom:00, yu:aeromag:23}. 
Particularly, to maximize potential beamforming gains, the use of massive MIMO has been actively explored in satellite communications \cite{you:jsac:20, angeletti:access:20}. 

Despite its advantages, however, massive MIMO faces several practical challenges. 
To be specific, beamforming generally needs channel state information (CSI) \cite{khammassi:survey:24}, which incurs substantial signaling overhead and quickly becomes outdated due to long propagation delays. Additionally, if satellites lack on-board processing capabilities, CSI must be relayed to a central gateway for precoder computation, leading to additional latency. For these reasons, the 3GPP non-terrestrial networks (NTN) standard \cite{3gpp:tr:38_821} specifies that GEO satellites use a fixed-beam approach, steering beams toward fixed points on Earth without adapting to instantaneous CSI. Consequently, existing MIMO beamforming methods that rely on real-time CSI \cite{chris:twc:15, khammassi:survey:24} may be unsuitable for GEO satellite systems with massive MIMO. 
Motivated by this, we conduct a fundamental analysis on the performance of the fixed-beam approach in GEO satellite communications. 

It is interesting to connect the fixed-beam approach to the concept of multi-user diversity, where fixed beamforming vectors are used and users with favorable channel conditions are selected. 
In the MIMO literature, this is also known as random or opportunistic beamforming \cite{sharif:tit:05,gilwon:twc:2016,zorba:space:08}.
Specifically, in \cite{sharif:tit:05}, the rate scaling law was analyzed according to the number of antennas and users in a rich scattering channel environment when a transmitter uses predetermined orthogonal beams. 
Assuming a single-path channel case, \cite{gilwon:twc:2016} explored the rate scaling law, capturing the interaction between the number of antennas and the users. 
There are two main limitations in applying these results to GEO satellite systems. First, the analysis is limited to a two-dimensional (2D) terrestrial network, which does not apprehend the three-dimensional (3D) geometry of GEO satellite systems. Second, each user is required to send feedback on the best signal-to-interference-plus-noise ratio (SINR) for user selection, leading to significant overheads. 
In \cite{koyoungcahi:twc:2022}, considering the 3D geometry, the outage probability and symbol error rate were analyzed for GEO satellite systems by modeling the locations of ground users as a homogeneous Poisson point process (PPP). 
However, a rigorous analysis of the rate scaling law in relation to the number of antennas and user density is missing in \cite{koyoungcahi:twc:2022}, limiting the understanding of the interplay between massive MIMO and the user density. To bridge this gap, in our previous work \cite{ksy:arxiv:2024}, the rate scaling law was derived for multiple unicast GEO satellite communications under a homogeneous PPP based user distribution.

{\textcolor{black}{In this paper, we extend the analytical framework in \cite{ksy:arxiv:2024} by considering a multicast scenario, which has been widely adopted in GEO satellite systems \cite{joroughi:twc:2017,shi:tbr:2023}. While these prior works, such as \cite{joroughi:twc:2017} exploiting SVD for robust precoder design and \cite{shi:tbr:2023} considering angle domain channels, offer advancements in multicast satellite communications, they still involve designing a precoder based on channel characteristics. In contrast, our work uniquely focuses on a practical fixed-beam transmission strategy combined with location-based user selection to serve multicast ground users. By modeling user locations as a PPP, we derive the asymptotic rate scaling law as a function of the number of antennas and the scaling factors of user density and multicast users.}}
Our contributions are as follows:
i) Our analysis shows that the rate scaling law is governed by the expression \( q -t - 1 \), where \( q \) and \( t \) denote the scaling factors of user density and multicast user count, respectively. This reveals that the rate degradation caused by increasing multicast users can be precisely compensated by a proportional increase in user density.
ii) We characterize the interference behavior as a function of the number of antennas, the number of multicast users, and the user density, thus enabling extension of our analysis to multibeam multicast satellite communications.
 

\section{System Model}
We consider a downlink GEO satellite equipped with a uniform planar array (UPA) consisting of \(M_{\text{x}} \times M_{\text{y}}\) antennas, while each multicast user is equipped with a single antenna.
We assume $M_x=M_y=M$, therefore the total number of antennas is $M^2$. 
The altitude of the GEO satellite is $H=$ 35786 km.
To serve $N$ multicast users, the GEO satellite forms a single beam. 
In Section \ref{sec:interference}, we extend this single beam configuration to a multibeam configuration.
We model the spatial locations of ground users with a homogeneous PPP $\Phi=\{\mathbf{x}_i \in \mathbb{R}^2, i \in \CMcal{N}\}$ with an intensity $\lambda$. 
The beam coverage region, denoted $\CMcal{A}$, is assumed to be a disk with radius $D$. 
The number of users included in the coverage region, denoted $|\CMcal{N}|$, follows the Poisson distribution with mean $\lambda |\CMcal{A}|= \lambda \pi D^2$.

1)~\emph{Channel Model}: 
Following a frequency-flat Ka-band channel assumption \cite{Gharanjik:icassp:2015}, 
we assume the channel for user $i$ as 
\begin{align} 
    \mathbf{h}_i = \sqrt{L_i} M \mathbf{v}_i \in \mathbb{C}^{M^2\times 1}.
\end{align}
$L_{i} = \left(\frac{c_0}{4\pi f_c d_i} \right)^2$ is the large-scale fading for user $i$ where $c_0$ and $f_c$ denote the speed of light and the carrier frequency. $d_i$ is the distance from the satellite to the user $i$. The array steering vector is defined as $\mathbf{v}_i \triangleq \mathbf{v} (\vartheta^x_i) \otimes \mathbf{v} (\vartheta^y_i)$ where 
\begin{align}
    \mathbf{v}(x) = \frac{1}{\sqrt{M}} \left[1, e^{-j\pi x}, \cdots, e^{-j\pi(M-1)x} \right]^{\sf T}. \nonumber
\end{align}
Here, $\otimes$ denotes the Kronecker product and
\begin{align}\label{eq:vartheta}
    \vartheta_{i}^x = \sin{\theta_{i}} \cos{\phi_{i}},~\vartheta_{i}^y = \sin{\theta_{i}} \sin{\phi_{i}},
\end{align}
where $\theta_i$ is the elevation angle and $\phi_i$ is the azimuth angle of the user $i$, respectively \cite{ksy:arxiv:2024}.

For simplicity, we omit rain attenuation since it does not affect the scaling law. This is because ground users within the same beam coverage experience the same rain attenuation \cite{ksy:arxiv:2024, koyoungcahi:twc:2022}. 
Since the altitude of a satellite is much greater than the distance of any reflection path to a user, the length of the reflection path can be ignored and the channel is assumed to be line-of-sight (LoS) \cite{ li:tcom:22, you:jsac:20,angeletti:access:20,zorba:space:08}.

{\textcolor{black}{
2)~\emph{User Selection}: 
The GEO satellite employs a nadir-pointing beamforming vector, $\mathbf{f}_1 = \mathbf{v}(0) \otimes \mathbf{v}(0)$, placing the beam center at $[\vartheta_{1}^x,\vartheta_{1}^y] = [0,0]$ (where $\theta_1$ and $\phi_1$ are the elevation and azimuth angles, respectively).}

{\textcolor{black}{
We propose selecting $N$ users within a predefined radius $R$ from this beam center. The set of selected users, $\CMcal{S}$, is formally expressed as:
\begin{align}
    \CMcal{S} = \left\{ i \in \CMcal{N} \, \middle| \, \|\mathbf{x}_i\| < R \right\}, \; |\CMcal{S}| = N, \label{eq:user selection}
\end{align}
where $R < D$. This radius $R$ is crucial as it implicitly determines the minimum achievable beam gain for selected users, directly influencing the multicast rate performance. While other criteria, such as proportional fairness, could be considered for selecting $N$ users within this radius, our primary objective in this paper is to clearly elucidate the fundamental relationship between beam gain and user distance in the asymptotic regime. Therefore, we adopt a random selection of $N$ users within $R$. This approach allows us to establish a baseline understanding, representing a worst-case scenario for selection that guarantees a certain minimum antenna gain as the number of antennas $M$ approaches infinity. This specific focus simplifies the analysis, highlighting the dominant factor in massive MIMO systems and providing intuitive insights into satellite beamforming.
The detailed characterization of $R$ and its implications for beam gain will be provided in the next section.
}}

Our user selection method in \eqref{eq:user selection} shares a foundational similarity with the approach in \cite{ksy:arxiv:2024} by exploiting user distance for fixed-beam precoding. However, a key difference lies in the transmission model: \cite{ksy:arxiv:2024} addressed a unicast scenario by selecting the closest user, whereas our work focuses on multicast, requiring $N$ users per beam.
Simply increasing the number of selected users fundamentally changes the statistical distribution of user distances \cite{haenggi:tit:2005}. This means the analysis from \cite{ksy:arxiv:2024} cannot directly derive a closed-form expression for the rate scaling law with our \eqref{eq:user selection} criteria in a multicast setting. Thus, our necessity for a new analytical approach to characterize multicast performance is a key contribution distinguishing our work. More details follow in the next section.

\begin{figure}[t]
    \centerline{\resizebox{1\columnwidth}{!}{\includegraphics{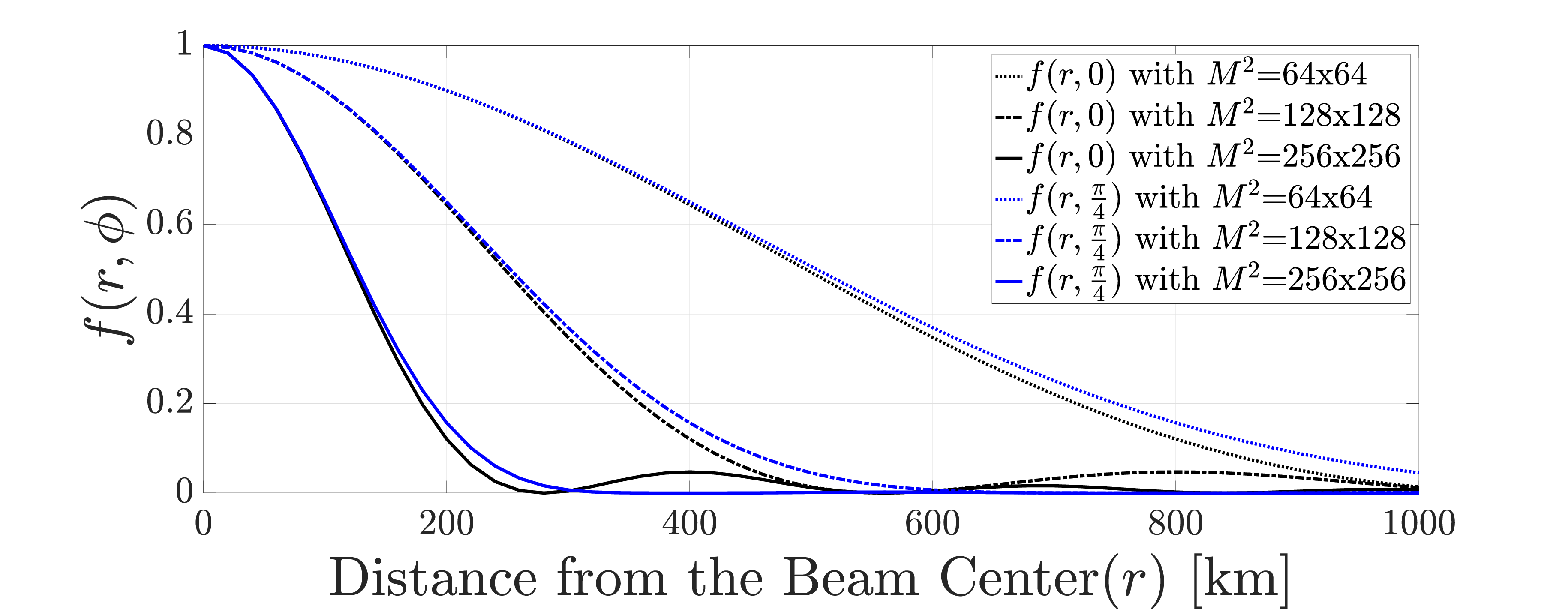}}} 
        \caption{Beam gain $f(r, \phi)$ for different values of $\phi$ with varying antenna array configurations $M^2 = 64 \times 64$, $M^2 = 128 \times 128$, and $M^2 = 256 \times 256$.}
        \label{fig:beam gain}
\end{figure}

\section{Scaling Law Analysis}
In this section, we analyze the scaling law of the ergodic rate. In the multicast scenario, the achievable rate is determined by the worst-case user \cite{koyoungcahi:twc:2022,joroughi:twc:2017} because the information rate should not be higher than the minimum achievable rate among $N$ multicast users. 
Assuming that users are at similar distances due to $H \gg r$, the received SNR is primarily determined by the distance $r$, as shown in Fig.~\ref{fig:beam gain}.
Based on this, we randomly select \( N \) users according to the user selection method described in \eqref{eq:user selection}, and without loss of generality, denote their indices such that \( \|{\bf{x}}_1\| \le \cdots \le \|{\bf{x}}_N\| \).
Consequently, the overall performance of the rate is determined by the user $N$. 
Denoting the beam gain with distance $r$ and azimuth angle $\phi$ as $f(r,\phi)$, 
the beam gain for user $N$, associated with the beamforming vector ${\bf{f}}_1$, is expressed using the $\text{Fej\'{e}r kernel}$ as 
\begin{align} 
    &f(r_N,\phi_N) 
    =  \left| \left[ \mathbf{v} (\vartheta_N^x) \otimes \mathbf{v} (\vartheta_N^y) \right]^{\sf H} \left[ \mathbf{v} (0) 
    \otimes \mathbf{v} (0) \right] \right|^2 \nonumber \\
    & \mathop{=} \frac{1}{M^4}  \left| \frac{ \sin\left(\frac{\pi M}{2} \sin\theta_N \cos\phi_N \right) \sin\left(\frac{\pi M}{2} \sin\theta_N \sin\phi_N \right)}
    { \sin\left(\frac{\pi }{2} \sin\theta_N \cos\phi_N \right) \sin\left(\frac{\pi }{2} \sin\theta_N \sin\phi_N \right)} \right|^2, \label{eq:beam_gain}
\end{align}
where $r_N = \|{\bf{x}}_N\|$ is the distance from user $N$ to the beam center. 
Then, the achievable rate of the user $N$ is given by
\begin{align}
    \mathcal{R}_N = \mathbb{E}\left[\log \left(1 + P  M^2 f (r_N, \phi_N)  \right) \right],
\end{align}
where $P$ is the SNR of user $N$. We define the array gain as $Z = M^2 f(r_N,\phi_N)$. 
To analyze $\mathcal{R}_N$ in an asymptotical regime, it is important to understand the asymptotical behavior of the array gain $Z$ for user $N$. To this end, we establish the following theorem. 
\begin{theorem}\label{thm:dist R}
    Assume that the multicast users are selected according to \eqref{eq:user selection} and denote the distance of the user $N$ as $r_N$. 
    In an asymptotic region where $M \rightarrow \infty$, if we choose $R$ such that 
    \begin{align}
            R \le \frac{H}{\sqrt{\frac{\pi^2}{4} M^{2p+\epsilon} -1 }}, \label{eq:radius}
        \end{align}
        then the probability of the event that $M^2 f(r,\phi)$ is larger than $M^{2p}$ is bounded as 
    \begin{align} \label{eq:Z bound}
            &\mathbb{P}\left[Z >M^{2p} \right]  \\
            &> \mathbb{P}\left[\frac{ H}{\sqrt{\alpha^2 M^{2+\epsilon/2}-1}} < r_N < \frac{ H }{\sqrt{4 \alpha^2 M^{(p+1) + \epsilon/2} -1 } } \right], \nonumber
    \end{align}
    where $\alpha = \frac{\pi}{4} \sqrt{|\sin\phi_N \cos\phi_N |}$ for large $M$.
\end{theorem}
\begin{proof}
Please refer to Appendix \ref{appendix: proof of Thm R}.
\end{proof}
Theorem \ref{thm:dist R} provides a key result on the range $R$ to achieve a certain beam gain in terms of the number of antennas, along with the corresponding probabilistic lower bound. 
By leveraging Theorem \ref{thm:dist R}, we derive the rate scaling law of the ergodic rate for multicast transmission. The following theorem is the main result of this paper. 
\begin{theorem} \label{thm:main}
    Let $N\sim M^t$ and $\lambda \sim M^q$ where $q \in (p+t+1+\epsilon/2, 2+t+\epsilon/2)$ and $t,p \in (0,1)$ with sufficiently small $\epsilon>0$. We derive the asymptotic upper and lower bounds of $\mathcal{R}_N$ as
    \begin{align}
        \log M^{2(q-t-1-\epsilon)} < \mathcal{R}_N < \log M^{2(q-t-1+\epsilon)}  \;\; \text{as $M\rightarrow \infty$}. \label{thm:scaling law}
    \end{align}
    Further, the fraction of $\mathcal{R}_N$ to the optimal unicast rate with perfect CSIT is derived as 
    \begin{align}
        \lim_{M \rightarrow \infty} \frac{ \mathcal{R}_N  }{ \mathbb{E}[ \log \left(1 +  P M^2 \right)]} = q-t-1.\label{thm: multicast N}
    \end{align}
\end{theorem}
\begin{proof}
To prove this, we first present the following lemma. 
\begin{lemma}\label{lem:dist of r}
The probability that the distance of user $N$, denoted $r_N$, is in the range between $R_a$ and $R_b$ is given by
\begin{align}
    \mathbb{P}\left[R_a < r_N < R_b\right] = \int^{\lambda \pi R_b^2}_{\lambda \pi R_a^2} \frac{x^{N-1}}{(N-1)!} e^{-x} dx.
\end{align}
\end{lemma}
\begin{proof}
    We can easily derive it by using $N$th neighbor distance such as
\begin{align}
    \mathbb{P}\left[R_a < r_N < R_b\right] & \mathop{=}^{\text{(a)} } \int_{R_a}^{R_b} \frac{2(\lambda \pi r^2)^N}{r\Gamma(N)}e^{-\lambda \pi r^2} dr, \notag \\
    &= \frac{\gamma(N, \lambda \pi R_b^2) - \gamma(N,\lambda \pi R_a^2) }{ \Gamma(N)}, \notag \\
    &= \int^{\lambda \pi R_b^2}_{\lambda \pi R_a^2} \frac{x^{N-1}}{(N-1)!} e^{-x} dx,
\end{align}
where (a) comes from the probability density function of $N$th closest point considering two-dimensional homogeneous PPP in \cite{haenggi:tit:2005}.
\end{proof}
By replacing the right-hand side of \eqref{eq:Z bound} with Lemma $\ref{lem:dist of r}$, we obtain the probability of the event $\{Z>M^{2p} \}$ such as
\begin{align}
    \mathbb{P}\left[Z >M^{2p} \right] &\mathop{>}^{\text{(a)}} \int^{\frac{\lambda \pi H^2}{4 \alpha^2 M^{(p+1) + \epsilon/2} -1} }_{\frac{\lambda \pi H^2}{\alpha^2 M^{2+\epsilon/2}-1}} \frac{x^{N-1}}{(N-1)!} e^{-x} dx, \label{eq:int gamma} \\
    & \mathop{\approx}^{\text{(b)}} \int^{U_N}_{L_N} \frac{1}{\sqrt{2\pi}}e^{-\frac{y^2}{2}}dy, \\
    &\mathop{\rightarrow}^{\text{(c)}} 1  \;\; \text{as $M\rightarrow \infty$}, \label{eq:prob 1}
\end{align}
where (a) is from Lemma \ref{lem:dist of r}. The PDF of $X \sim \text{Gamma}(k,\theta)$ with shape parameter $k$ and scale parameter $\theta$ is given by
\begin{align}
    f_X(x;k,\theta) = \frac{1}{\Gamma(k) \theta^k} x^{k-1} e^{-x/\theta},
\end{align}
which is defined on the interval $(0,\infty)$ and the integrand of \eqref{eq:int gamma} is corresponding to $f_X(x;N,1)$. 
(b) holds where for large $N$, $\text{Gamma}(k,\theta)$ converges to the normal distribution $\mathcal{N}(\mu,\sigma^2)$ where $\mu=k\theta$ and $\sigma^2 = k\theta^2$.
Then, we have $X\sim \mathcal{N}(N,N)$ which is easily converted to standard normal distribution by letting $Y = (X-N)/\sqrt{N}$ where
\begin{align}
    U_N = \frac{\frac{\lambda \pi H^2}{4 \alpha^2 M^{(p+1) + \epsilon/2} -1} - M^t }{M^{t/2}}, \; 
    L_N = \frac{\frac{\lambda \pi H^2}{\alpha^2 M^{2+\epsilon/2}-1}- M^t}{M^{t/2}}.
\end{align}
Under condition $q \in (p+t+1+\epsilon/2,2+t+\epsilon/2)$, (c) holds because $U_N \rightarrow \infty$ and $L_N \rightarrow -\infty$ and the integration of any PDF throughout the region is always $1$.

We know that $\mathbb{P}[Z>M^{2p}] \rightarrow 1$ for $q \in (p+t+1+\epsilon/2,2+t+\epsilon/2)$. Following the proof technique in \cite{gilwon:twc:2016}, we set $p = q-t-1-\epsilon$. Then, the lower bound of $\mathcal{R}_N$ is given by
\begin{align}
    \mathcal{R}_N &\ge \int^{M^2}_{M^{2(q-t-1-\epsilon)}} \log(1+Pz)p(z)dz, \notag \\
    &\ge \log(1+PM^{2(q-t-1-\epsilon)})  \int^{M^2}_{M^{2(q-t-1-\epsilon)}} p(z)dz, \notag \\
    &\mathop{\rightarrow} 2(q-t-1-\epsilon) \log M + \log P \;\; \text{as $M\rightarrow \infty$}.
\end{align}
By setting $p=q-t-1+\epsilon$, the upper bound is given by
\begin{align}
    \mathcal{R}_N &= \int^{M^2}_{M^{2(q-t-1+\epsilon)}} \log(1+Pz) p(z) dz, \notag \\
    &\phantom{-----} +\int^{M^{2(q-t-1+\epsilon)}}_{0} \log(1+Pz) p(z) dz, \notag \\
    &\mathop{\le}^{\text{(a)}} \log(1+PM^{2(q-t-1+\epsilon)} ) \int^{M^{2(q-t-1+\epsilon)}}_{0}  p(z) dz, \notag\\
    &\rightarrow 2(q-t-1+\epsilon) \log M + \log P \;\; \text{as $M\rightarrow \infty$},
\end{align}
where (a) is from the fact that $\mathbb{P}[Z>M^{2p}]\rightarrow 0$ for $q<p+t+1+\epsilon$. 
Letting $M \rightarrow \infty$, this completes the proof of \eqref{thm:scaling law}. 
To derive \eqref{thm: multicast N}, we characterize the bounds of $\mathcal{R}_N$:
\begin{align}
    \frac{2(q-t-1-\epsilon) \log M + \log P}{2\log M + \log P} \le \frac{\mathcal{R}_N }{\log(1+PM^2)} \\
    \phantom{------} \le \frac{2(q-t-1+\epsilon) \log M + \log P}{2\log M + \log P}. \notag
\end{align}
As $M \rightarrow \infty$ for small $\epsilon>0$, we complete the proof of \eqref{thm: multicast N}.
\end{proof}

\begin{figure}[t]
    \centerline{\resizebox{1\columnwidth}{!}{\includegraphics{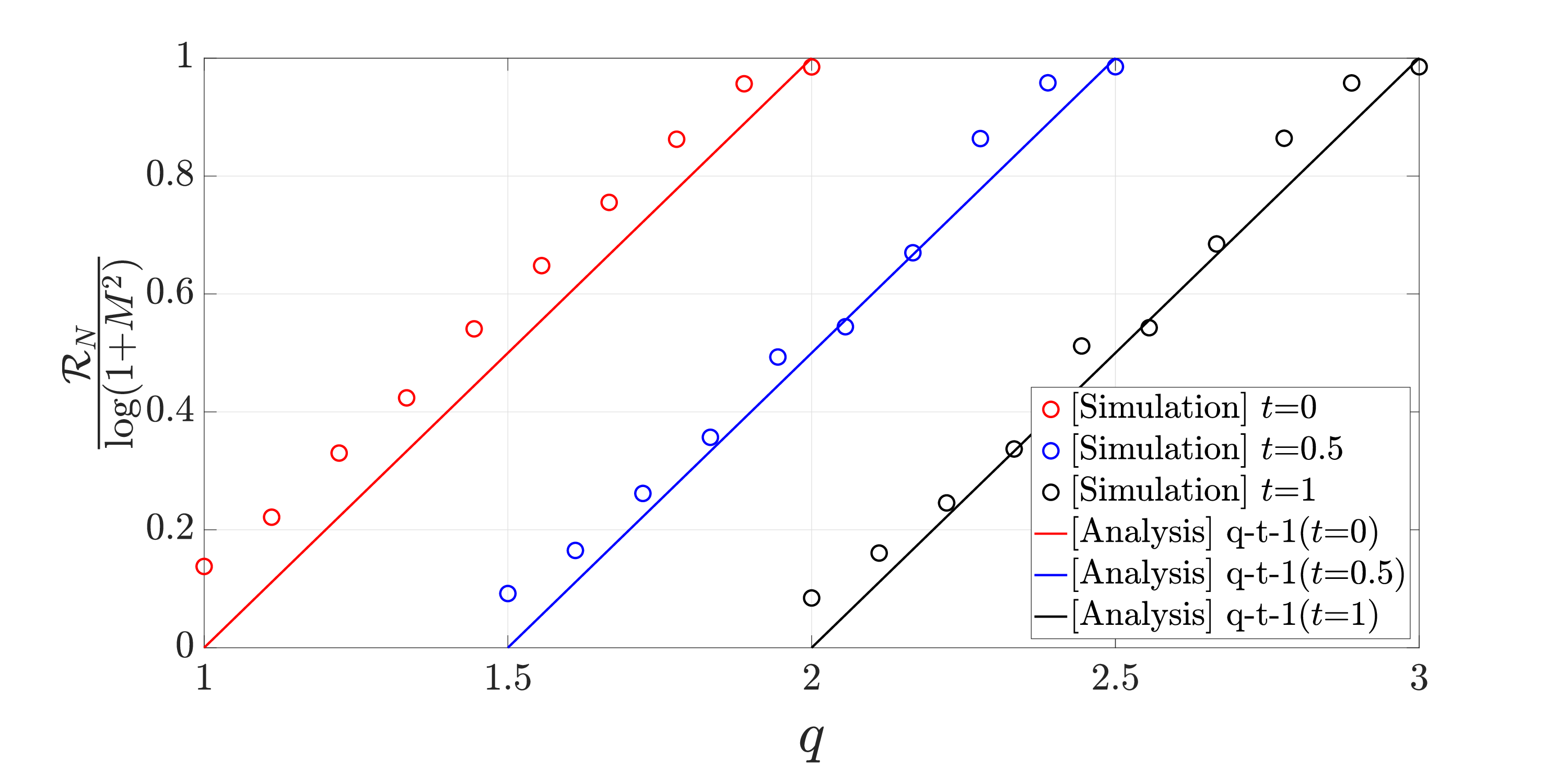}}} 
        \caption{The ratio of $\mathcal{R}_N$ and ideal rate $\log(1+M^2)$ versus $q$ for different $t$ where $M=10^4$.}
        \label{fig:multicast N}
\end{figure}
Theorem \ref{thm:main} offers valuable insights into GEO satellite communications with massive MIMO. Specifically, \eqref{thm:scaling law} characterizes user density, multicast users, and antenna scaling. If multicast users $N$ are constant, it does not affect the rate scaling law. However, if $N \sim M^t$, achievable rate scaling diminishes, as expected. Moreover, increased user density ($\lambda \sim M^q$) enhances rate scaling by $q$, due to higher probability of selecting $N$ multicast users within $R$. Notably, $q$ and $t$ factors precisely offset, yielding asymptotic rate scaling $\mathcal{R}_N \sim (q-t-1)\log M^{2}$ for small $\epsilon$. This means fixed-beam precoding asymptotically achieves a fraction $q-t-1$ of the ideal rate with perfect CSI.

%
%
%

To validate the analytical insights provided by Theorem \ref{thm:main}, numerical evaluations illustrating the interplay between user density and the number of multicast users are presented in Fig.~\ref{fig:multicast N}. In Fig.~\ref{fig:multicast N}, $\mathcal{R}_N$ represents the ergodic rate of the $N$th user and $\log(1+M^2)$ corresponds to the ideal rate, achieved by aligning the beamforming vector with the user channel under the assumption of perfect CSI. Assuming $P=1$ for simplicity, Fig.~\ref{fig:multicast N} shows the fraction of $\mathcal{R}_N$ and the ideal rate versus $q$ for various $t$ (0, 0.5, 1). These results confirm multicast transmission with fixed-beam precoding asymptotically achieves the $q-t-1$ fraction of optimal ergodic rate attainable by dynamic beamforming with perfect CSI.\footnote{{\textcolor{black}{Although theoretical convergence of the scaling law typically assumes large $M$, our numerical results demonstrate practical convergence for moderate $M$ values (e.g., $M>2048$).}}}
This highlights that fixed-beam precoding multicast necessitates additional user density compared to unicast scenarios \cite{ksy:arxiv:2024}. 


{\textcolor{black}{
    To do this, we compare the multicast characteristics presented in this work with the unicast case analyzed in \cite{ksy:arxiv:2024}. Specifically, let $\mathcal{R}_1$ denote the ergodic rate of unicast transmission under the fixed-beam approach. Then, for $\lambda \sim M^q$ with $q\in(p+1+\epsilon/2, 2+\epsilon/2)$ for $p\in(0,1)$ with arbitrarily small $\epsilon>0$, we have
    \begin{align}
        \lim_{M \rightarrow \infty} \frac{\mathcal{R}_1}{\mathbb{E}[\log(1+PM^2)]} = q-1. \label{eq: unicast frac}
    \end{align} 
    Comparing the multicast scaling law in Theorem \ref{thm:main} with the unicast case in \eqref{eq: unicast frac}, we observe that multicast transmission inherently reduces the achievable fraction by a factor of $t$.
    To compensate for this reduction, an additional user density scaling $\lambda \sim M^{q+t}$ exactly offsets the performance loss incurred by multicast transmission. 
Specifically, adopting the stronger scaling $\lambda \sim M^{2+t}$ allows the achievable multicast rate to asymptotically match the unicast rate. These insights highlight the key trade-off between multicast group size and user density, emphasizing user density scaling's role in mitigating multicast rate reduction.
}}

\section{Interference Analysis} \label{sec:interference}
In this section, we extend our framework to multibeam multicast satellite communications that employ $M^{2\ell}$ beams.
In this multibeam scenario, the coverage area is partitioned into a uniform grid, with each fixed beam directed toward the center of a grid element. The configuration of the kth fixed beam is given by:
    $\vartheta^x_k = \sin \theta_k \cos \phi_k = \frac{2n}{M^\ell}, \; 
    \vartheta^y_k = \sin \theta_k \cos \phi_k = \frac{2n}{M^\ell}$
where $\{n,m\} \in \mathbb{Z}$. $\theta_k$ and $\phi_k$ denote the elevation and azimuth angle of $k$th beam, respectively, and $\ell$ controls the beam spacing. 
With this setup, the beamforming vector for the $k$th beam is expressed as $\mathbf{f}_k = \mathbf{v}(\vartheta_k^x) \otimes \mathbf{v}(\vartheta_k^y)$. 
The interference received from an adjacent beam $j$ is quantified by $\frac{M^2}{M^{2\ell}} f_j(r_N,\phi_N)$, where the factor $M^{2\ell}$ arises from the power adjustment due to multiple beams. This leads us to the following theorem quantifying interference.
{\textcolor{black}{
\begin{theorem}\label{thm:interference}
    For $s,\ell \in (0,1)$ such that $s + \ell <1$ as $M\to\infty$, we have
    \begin{align}
        \mathbb{P}\left[ \frac{M^2}{M^{2\ell}} f_j(r_N,\phi_N) < \frac{1}{M^{2s}} \right] &> \int^{\frac{\lambda \pi H^2}{M^{2\ell}-1}}_0 \frac{x^{N-1}}{(N-1)!}e^{-x} dx, \nonumber \\
        &= \frac{x^N}{N!} {}_1F_1\left(N,N+1;-x\right) \bigl|_{x = \frac{\lambda \pi H^2}{M^{2\ell}-1}}. \nonumber
    \end{align}
\end{theorem}
}}
\begin{proof}
A brief outline is provided here. We begin by splitting the condition $\frac{M^2}{M^{2\ell}} f_j(r_N,\phi_N) < \frac{1}{M^{2s}}$ into two sufficient conditions, yielding a lower bound on the probability. Applying the techniques detailed in Theorem \ref{thm:main}, we derive the probability term given on the right-hand side. For comprehensive derivation steps, readers are referred to \cite{ksy:arxiv:2024}, with additional results from Lemma \ref{lem:dist of r} of this paper.
\end{proof}

{\textcolor{black}{Theorem \ref{thm:interference} highlights the relationship between the number of beams and inter-beam interference levels. While the expression might initially seem complex, it is essentially an incomplete Gamma function, widely recognized and available in standard mathematical libraries, making it a compact and efficiently computable representation for practical use. The tightness of this derived bound has been rigorously confirmed through extensive comparisons with simulation results. These comparisons consistently show strong concordance between our analytical calculations (from Theorem \ref{thm:interference}) and direct interference measurements from PPP-distributed users, validating that our analysis accurately captures the multicast interference probability across various system parameters $q$ and $t$.}}


\section{Conclusion}
In this paper, we analyzed the rate scaling law of multicast transmission in a GEO satellite system with massive MIMO using a fixed-beam approach. Assuming ground users are distributed according to a homogeneous PPP, we derived the asymptotic rate scaling law as a function of the number of antennas and the scaling factors of user density and multicast users. Our analysis revealed that the rate scaling law follows the expression \( q - t - 1 \), where \( q \) and \( t \) represent the scaling factors of user density and multicast user count, respectively. Consequently, we demonstrated that the performance degradation due to an increasing number of multicast users can be exactly compensated by proportionally increasing the user density. Furthermore, by characterizing interference in relation to the number of antennas, multicast users and user density, our findings provide foundational insights for extending this framework to multibeam multicast satellite communication systems.

\appendices

\section{Proof of Theorem \ref{thm:dist R}} \label{appendix: proof of Thm R}
We divide the event $\{Z>M^{2p} \}$ into two cases: $\phi=0$ and $\phi \neq 0$.
The sufficient condition for the event $\{Z>M^{2p}|\phi \neq 0 \}$ is given by
\begin{align} \label{eq:suff1}
    \left|  \sin\left(\frac{\pi M }{2} \sin\theta_N \cos\phi_N  \right) \sin\left(\frac{\pi M }{2} \sin\theta_N \sin\phi_N \right) \right| > \frac{1}{M^{\epsilon/2}},
\end{align}
and
\begin{align}\label{eq:suff2}
    \left| \sin\left(\frac{\pi }{2} \sin\theta_N \cos\phi_N \right) \sin\left(\frac{\pi }{2} \sin\theta_N \sin\phi_N \right) \right| < \frac{1}{M^{(p+1)+\epsilon/2}}.
\end{align}
with sufficiently small $\epsilon>0$. From the fact $| \sin x | > \frac{|x|}{2}$, the sufficient condition $\eqref{eq:suff1}$ is given by
$    \frac{1}{2} \left| \frac{\pi M}{2} \sin\theta_N \cos\phi_N \right| \cdot \frac{1}{2}\left| \frac{\pi M}{2} \sin\theta_N \sin\phi_N \right| > \frac{1}{M^{\epsilon/2}}$, 
which leads to a lower bound 
\begin{align}
    | \sin \theta_N | > \frac{1}{ \frac{\pi}{4} \sqrt{|\sin\phi_N \cos\phi_N|} M^{1+\epsilon/4}}. \notag
\end{align}
From the sufficient condition \eqref{eq:suff2} with the fact that $|\sin x| < |x|$ for $x\in(0,\pi/2)$, we derive 
    $\left|\frac{\pi}{2} \sin\theta_N \cos\phi_N \right| \cdot \left| \frac{\pi}{2} \sin\theta_N \sin\phi_N \right| < \frac{1}{M^{(p+1)+\epsilon/2}} $
and finally obtain an upper bound
\begin{align}
    |\sin \theta_N| < \frac{1}{\frac{\pi}{2} \sqrt{|\sin\phi_N \cos\phi_N|} M^{(p+1)/2 + \epsilon/4}}. \notag
\end{align}
By combining the results, we can obtain the sufficient condition for the event $\{Z>M^{2p} |\phi\neq 0 \}$ as
\begin{align} \label{eq:range sin theta}
    \frac{1}{ \alpha M^{1+\epsilon/4}} < |\sin\theta_N| < \frac{1}{2\alpha  M^{(p+1)/2 + \epsilon/4}}
\end{align}
where $\alpha = \frac{\pi}{4} \sqrt{|\sin\phi_N \cos\phi_N |}$ is the constant for given $\phi$. 
Then, by using $|\sin \theta_N| = r_N / \sqrt{r_N^2 +H^2}$, we have
\begin{align}
    &\mathbb{P}\left[Z >M^{2p} \right] \nonumber \\
    &> \mathbb{P}\left[\frac{ H}{\sqrt{\alpha^2 M^{2+\epsilon/2}-1}} < r_N < \frac{ H }{\sqrt{4 \alpha^2 M^{(p+1) + \epsilon/2} -1 } } \right].
\end{align}
Here, we derive the case $\phi =0$. The event $\{Z>M^{2p} \}$ is given by
\begin{align} 
    \frac{1}{M^2} \left| \frac{ \sin\left(\frac{\pi M}{2} \sin\theta_N \right) }
    { \sin\left(\frac{\pi }{2} \sin\theta_N  \right) } \right|^2 > M^{2p}. \label{eq:M>Z, phi =0}
\end{align}
From this, we obtain the sufficient condition for \eqref{eq:M>Z, phi =0}
\begin{align}
    \left| \sin\left(\frac{\pi M}{2} \sin \theta_N \right) \right| > \frac{1}{M^{\epsilon/4}} \text{  $\&$  } \left| \sin\left(\frac{\pi }{2} \sin \theta_N \right) \right| < \frac{1}{M^{p+ \epsilon/4}}. \notag
\end{align}
With an approach similar to the case $\phi \neq 0$, we can easily derive the sufficient condition for $\{ Z>M^{2p} | \phi = 0 \}$ as following:
\begin{align}
    \frac{H}{\sqrt{\frac{\pi^2}{16}M^{2+\epsilon/2} -1 } } < r_N < \frac{H}{\sqrt{\frac{\pi^2}{4} M^{2p + \epsilon/2} -1}}.
\end{align}
We have $\frac{ H}{ \sqrt{4 \alpha^2 M^{(p+1) + \epsilon/2} -1 } } < \frac{H}{\sqrt{\frac{\pi^2}{4} M^{2p + \epsilon/2} -1}},$ for sufficiently large $M$.
It is obvious that for $\theta_N=0$, i.e., $r_N=0$, $Z$ has maximum value $M^2$. 
In $\text{Fej\'{e}r kernel}$, the point of $\theta_N =0$ becomes a singularity point, which cannot be evaluated. 
This makes a lower bound of the probability for $\mathbb{P}[Z>M^{2p}]$ in \eqref{eq:Z bound}.
This completes the proof. 

\bibliographystyle{IEEEtran}
\bibliography{Multicast.bib}

\end{document}